\documentstyle[sprocl]{article}

\begin{document}

\title{KN interaction in the quark-gluon exchange framework}
\author{Dimiter Hadjimichef}
\address{Instituto de F\'{\i}sica e Matem\'{a}tica, UFPel, Caixa Postal 354\\
96010-900, Pelotas, R.S.,Brazil.\\ E-mail: dimiter@ufpel.tche.br}
\maketitle
\abstract The general meson-baryon potential obtained in the context of the
\ Fock-Tani formalism is applied in the $K^{+}N$ system. A simple
spin-dependent interaction on quark level is used in order to compare with
known results. The short range nature of the equivalent local potential is
shown.

\section{Introduction}

One of the most extensively studied interactions is the $NN$ interaction. In
the context of the Bonn potential the short range repulsion arises from an $%
\omega $-meson exchange with a larger coupling constant
than obtained from $SU(6)$ symmetry estimates \cite{holinde1}. In the $KN$-Bonn
potential the $K^{+}N$ repulsion also arises from an $\omega $ exchange. If
one keeps the $SU(6)$ coupling, a very short ranged phenomenological meson
must be added. When one switches to the $K^{-}N$ system, the $\omega$ 
becomes attractive. In order to fit all the partial-waves, the $\omega$ coupling constant must
be assigned the $SU(6)$ value  and the additional repulsive contribution reduced
by a factor of five \cite{kn}. It whould interesting to investigate if this can be explained in 
the context of meson exchange or whether it is an effect of the quark-gluon dynamics.
 In the constituent quark model the $K^{+}=u\,\bar{s}$ and $N=uud$ or $udd$ so the 
simple quark exchange mechanism can be applied to the $K^{+}N$ system \cite{kn-many}. 
A  simple spin-spin interaction on quark level is used in order to compare with
a similar calculation \cite{barnes}.

\section{The Kaon-Nucleon Potential}
The method we employ in order to introduce the quark-gluon degrees of
freedom is known as the Fock-Tani formalism \cite{ft}.   The meson-baryon
potential has been obtained in a standard way \cite{dimi99}:
\begin{equation}
V_{{\rm meson-baryon}}=\sum_{i=1}^4\;V_i(\alpha \beta ;\delta \gamma )\;{{%
m_\alpha ^{\dagger }\,}b_\beta ^{\dagger }\,m_\gamma \,b_\delta }
\label{vkn_1}
\end{equation}
\newpage
\noindent and  
\begin{eqnarray}
V_1{(\alpha \beta ;\delta \gamma )} &=&-3V_{qq}(\mu \nu ;\sigma \rho )\,\Phi
_\alpha ^{*\mu \nu _2}\Psi _\beta ^{*\nu \mu _2\mu _3}\Phi _\gamma ^{\rho
\nu _2}\Psi _\delta ^{\sigma \mu _2\mu _3}\;\   \label{vkn_2} \nonumber\\
V{_2{(\alpha \beta ;\delta \gamma )}} &=&{-3V_{q\overline{q}}(\mu \nu
;\sigma \rho )\,\Phi _\alpha ^{*\mu _1\nu }\Psi _\beta ^{*\mu \mu _2\mu
_3}\Phi _\gamma ^{\sigma \rho }\Psi _\delta ^{\mu _1\mu _2\mu _3}}  \nonumber
\\
V{_3{(\alpha \beta ;\delta \gamma )}} &=&{-3V_{qq}(\mu \nu ;\sigma \rho
)\,\Phi _\alpha ^{*\mu \nu _2}\Psi _\beta ^{*\mu _1\nu \mu _3}\Phi _\gamma
^{\mu _1\nu _2}\Psi _\delta ^{\sigma \rho \mu _3}}  \nonumber \\
V{_4{(\alpha \beta ;\delta \gamma )}} &=&{-6V_{q\overline{q}}(\mu \nu
;\sigma \rho )\,\Phi _\alpha ^{*\nu _1\nu }\Psi _\beta ^{*\mu _1\mu \mu
_3}\Phi _\gamma ^{\mu _1\rho }\Psi _\delta ^{\nu _1\sigma \mu _3}.} 
\end{eqnarray}
The general potential (\ref{vkn_1}) can specialized to study the KN system. 
The microscopic quark-quark potential and quark-antiquark can be written as 
\cite{barnes}
\[
V_{qq\;{\rm or}\;\overline{qq}}=\sum_{a,i<j}\;\left( -\frac{8\pi \alpha _s}{%
3m_{i\,}m_j}\right) \;\left[ \vec{S}_i\,\cdot \vec{S}_j\right] \,\left[ 
{\cal F}_i^a\cdot \,{\cal F}_j^a \right], \;\;\ 
\]
where $S_{i}$, ${\cal F}_{i}$ are the spin and color matrixes of the quarks. 
The meson and baryon wave functions are written as a product of the
quark color-spin-isospin-space wave functions  \cite{ft}
\begin{eqnarray*}
&&\Phi_\alpha ^{\mu \nu } =\delta (\vec{p}_\alpha -\vec{p}_\mu -\vec{p}%
_\nu )\;\frac{\delta^{c_\mu c_\nu }}{\sqrt{3}}\,\;\frac{\chi _\alpha
^{\,f_\mu \,f_\nu }}{\sqrt{2}}\,\;\varphi (\vec{p}_\mu ,\vec{p}_\nu ) \\
&&\Psi _\alpha ^{\mu _1\mu _2\mu _3} =\delta (\vec{p}_\alpha -\vec{p}
_{1}-\vec{p}_{2}\vec{p}_{3})\,N(p_\alpha )\,\frac{%
\varepsilon ^{c_1c_2c_3}}{\sqrt{6}}\frac{\chi_{\alpha} ^{^{f_1f_2f_3}}}{\sqrt{18}}%
\phi (\vec{p}_{1})\phi (\vec{p}_{2})\phi (\vec{p}_{3}).
\end{eqnarray*}
The spatial part of the meson wave function is chosen to be a gaussian
\begin{eqnarray*}
\varphi (\vec{p}_{q},\vec{p}_{\bar{q}})=\left( \pi \beta ^2\right) ^{-\frac 34}\,\exp
\left[ -\frac{\left( m_1\vec{p}_{q} - m_2\vec{p}_{\bar{q} }   \right) ^2}{8\,\beta ^2}%
\right] 
\end{eqnarray*}
with $m_1=\frac{2\,m_{\bar{q}}}{m_q+m_{\bar{q}}}$ and $m_2=\frac{2\,m_q}{m_q+m_{\bar{q}}}$.
The spatial part of the baryon wave function is also written in terms of a gaussian
\begin{eqnarray*}
\phi (\,\vec{p}\,) = \left(\frac{b^2}{\pi}\right)^{\frac 34}\,\exp \left[ -\frac
12\,b^2p^2\right]\,\,\,\, ,\, \,\,\,
N(\,\vec{p}\,) = \left(\frac{3\pi }{b^2}\right) ^{\frac 34}\,\exp \left[ \frac
16\,b^2p^2\right].
\end{eqnarray*}
In the CM coordinate system $\vec{p}_\alpha =-\vec{p}_\beta =\vec{p}$ ,
$\vec{p}_\gamma =-\vec{p}_\delta =\vec{p}\;^{\prime }$ 
and using the notation of\, \cite{barnes}: $m_1=\frac 2{1+\rho }$ ,
$m_2=\frac{2\,\rho }{1+\rho }$ ,
$b=1/ \alpha $ , $\beta =\alpha/\sqrt{g}$ ,
where $\rho =m_q/m_{\bar{q}},$ one finds for the kernel of the potential 
(\ref{vkn_1}), apart from an overall energy conservation $\delta$, the spatial
part is given by
\begin{eqnarray}
V_1(\vec{p},\vec{p}^{\prime }) &=&3\;k_{ss}
\eta _1\;\exp \left[ -\,A\,\left( \,\vec{p}-\vec{%
p}^{\prime }\,\right) ^2\right]  \nonumber \\
V_2(\vec{p},\vec{p}^{\prime }) &=&3\;k_{ss} \eta _2\;\exp \left[
-\,B_{1\;}p^2-B_2\;p^{\prime \,2}\,+\ B_3\;\,\vec{p}\cdot \vec{p}^{\prime
}\right]  \nonumber \\
V_3(\vec{p},\vec{p}^{\prime }) &=&3\;k_{ss}
\eta _3\;\exp \left[
-\,C_{1\;}p^2-C_2\;p^{\prime \,2}\,+\ C_3\;\,\vec{p}\cdot \vec{p}^{\prime }\
\right]  \nonumber \\
V_4(\vec{p},\vec{p}^{\prime }) &=&6\;k_{ss}
\eta _4\;\exp \left[ -\,D_1\,\left( \,p^2\
+p^{\prime \,2}\right) +D_2\;\,\vec{p}\cdot \vec{p}^{\prime }\right] 
\label{vkn_4}
\end{eqnarray}
with 
\begin{eqnarray*}
\eta _1 &=&1\;\;;\;\;\eta _2=\rho \,\left( \frac{6}{g+3}\right) ^{3/2} \;\;;\;\;
\eta _3 =\left(\frac {12g}{7g+6}\right) ^{3/2}\\
\eta _4&=&\rho \,\left(\frac
{12g}{\ (2g+3)(g+2)}\right) ^{3/2} \;\;\;\; ; \;\;\;\;
k_{ss}= \frac{8\pi \alpha _s}{3m_q^2 (2\pi )^3}
\end{eqnarray*}
and 
\begin{eqnarray*}
A &=&\frac{2(1+\rho )^2+3g}{12\alpha ^2(1+\rho )^2} \\
B_1 &=&\frac{(5g+3)\rho ^2+(-2g+6)\rho +(2g+3)}{6\alpha ^2(g+3)(1+\rho )^2}
\\
B_2 &=&\frac{(5g+3)\rho ^2+(10g+6)\rho +(5g+3)}{6\alpha ^2(g+3)(1+\rho )^2}
\\
B_3 &=&\frac{(-g+1)\rho ^2+2\rho +(g+1)}{\alpha ^2(g+3)(1+\rho )^2} \\
C_1 &=&\frac{(16g+3)\rho ^2+(\ 2g+6)\rho +(21g^2+22g+3)}{12\alpha
^2(7g+6)(1+\rho )^2} \\
C_2 &=&\frac{(64g+3)\rho ^2+(\ 62g+6)\rho +(21g^2+34g+3)}{12\alpha
^2(7g+6)(1+\rho )^2} \\
C_3 &=&\frac{(-8g+1)\rho ^2+2\rho +\ (7g^2+8g+1)}{2\alpha ^2(7g+6)(1+\rho )^2%
} \\
D_1 &=&\frac{(20g^2+40g+3)\rho ^2+\ (4g^2+14g+6)\rho +\ (5g^2+10g+3)}{%
12\alpha ^2(2g+3)(g+2)(1+\rho )^2} \\
D_2 &=&\frac{(-4g^2-8g+1)\rho ^2+\ (-4g^2-6g+2)\rho +\ (\ g^2+2g+1)}{\
2\alpha ^2(2g+3)(g+2)(1+\rho )^2 }\;\;.
\end{eqnarray*}
In order to obtain a potential in the configuration space one first
introduces the new variables
${\cal \vec{P}}=\frac 12(\vec{p}+\vec{p}^{\prime })$ and  
${\cal\vec{Q}}=(\vec{p}-\vec{p}^{\prime })$
in (\ref{vkn_4}) and uses the local approximation which
consists in ${\cal \vec{P}}=0.$
The configuration space $V(r)$ is a simple Fourier transform of
(3) times the color-spin-isospin factor $\omega_{i}$ 
\begin{eqnarray*}
V(r)=\sum_{i=1}^4\int d{\cal Q\;}e^{i{\cal \vec{Q}}\cdot \,\vec{r}}
\omega_{i}\; V_i({\cal Q})=k_{ss}\sum_{i=1}^4V_i(r) 
\end{eqnarray*}
where  
\begin{eqnarray*}
V_1(r) &=&\left(\frac {\pi }{A}\right) ^{3/2}\omega _{1\,}\eta _1\exp \left( -\frac{%
r^2}{4A}\right) \\
V_2(r) &=&\left(\frac {4\pi }{B_1+B_2+B_3}\right) ^{3/2}\omega _{2\,}
\eta _2\exp
\left( -\frac{r^2}{\ B_1+B_2+B_3}\right) \\
V_3(r) &=&\left(\frac {4\pi }{\ C_1+C_2+C_3}\right) ^{3/2}\omega _3\,\eta _3\exp
\left( -\frac{r^2}{\ C_1+C_2+C_3}\right) \\
V_4(r) &=&\left(\frac {4\pi }{2D_1+D_2}\right) ^{3/2}\omega _4\,\eta _4\exp \left( -%
\frac{r^2}{\ 2D_1+D_2}\right) .
\end{eqnarray*}
The color-spin-isospin factors $\omega _i$ are for $I=0$,  
$[0,\frac{1}{6},0,\frac{1}{6}]$ 
and $I=1$, $[\frac{1}{3},\frac {1}{18},\frac{1}{3},\frac {1}{18}]$. 
A graph for this potential can is seen in
figure 1 for the following set of parameters \cite{barnes}: $\alpha _s=0.6,$ $m_q=330$ MeV, 
$\rho =0.6,$ $\alpha =400$ MeV and $\beta =350$ MeV. 
\section{Conclusions}
The potentials are repulsive and short ranged, as one would expect for the 
nuclear hard core. However, in order to judge the adequacy of the quark-gluon exchange
hypothesis, one has to see whether the model provides, for example, the correct 
spin-orbit interaction in the P-waves. This requires a more elaborated calculation, in 
the context of a meson exchange model with the short distance interaction described
by the Fock-Tani KN potential.
\section*{Acknowledgements}
The author was partially supported by FAPERGS, IFM-UFPel.

\end{document}